\pdfoutput=1

\documentclass[10pt,journal,compsoc]{IEEEtran}
\ifCLASSOPTIONcompsoc
  \usepackage{cite}
\else
  \usepackage{cite}
\fi
\ifCLASSINFOpdf
\else
\fi
\usepackage[cmex10]{amsmath}
\ifCLASSOPTIONcompsoc
  \usepackage[caption=false,font=footnotesize,labelfont=sf,textfont=sf]{subfig}
\else
  \usepackage[caption=false,font=footnotesize]{subfig}
\fi
\usepackage{fixltx2e}
\usepackage{url}

\usepackage{amssymb}
\usepackage{mathtools}
\usepackage{graphicx}
\usepackage{fullpage}
\usepackage{booktabs}
\usepackage{xr}
\newcommand{\kay}{\langle k \rangle}
\DeclareMathOperator{\e}{e}
\allowdisplaybreaks

\begin{document}
%
\title{Design of Self-Organizing Networks: Creating specified degree distributions}
%
%
%
%

\author{Holly~Silk,~Martin~Homer~and~Thilo~Gross
\IEEEcompsocitemizethanks{
\IEEEcompsocthanksitem The authors are with the University of Bristol, Department of Engineering Mathematics and Bristol Centre for Complexity Sciences, Bristol, UK.\protect\\
E-mail: holly.silk@bristol.ac.uk}
}

\IEEEtitleabstractindextext{%
\begin{abstract}
A key problem in the study and design of complex systems is the apparent disconnection between the microscopic and the macroscopic.
It is not straightforward to identify the local interactions that give rise to an observed global phenomenon, nor is it simple to design a system that will exhibit some desired global property using only local knowledge. 
Here we propose a methodology that allows for the identification of local interactions that give rise to a desired global property of a network, the degree distribution.
Given a set of observable processes acting on a network, we determine the conditions that must satisfied to generate a desired steady-state degree distribution.
We thereby provide a simple example for a class of tasks where a system can be designed to self-organize to a given state.
\end{abstract}

}

\maketitle

\IEEEdisplaynontitleabstractindextext

%
\IEEEpeerreviewmaketitle


%
%
%
%
\IEEEraisesectionheading{\section{Introduction}\label{sec:introduction}}
\IEEEPARstart{C}{omplex} systems can exhibit phenomena and properties that are not inherent in the system's constituents but arise from their interactions. In particular, ordered structures can be formed without requiring pre-appointed hubs or leaders \cite{braha2006}. 

In biology the ability of complex systems to form macroscopic structures and patterns based on simple local rules is evident in all organisms and on all levels of organization. Examples range from the formation of complex (bio)molecules from simple chemical reactions, via the development of tissues and organisms, to social organization and collective decision-making \cite{redner2001}.

Technical systems too provide many examples of self-organization, including particular types of power-cuts \cite{carreras2004power}, traffic jams \cite{chowdhury1999traffic}, and structural instabilities in constructions \cite{eckhardt2007bridge}. While in biology self-organization is thus essential for the function, it often appears in technical systems primarily as a source of failure. 

The ability of biological systems to exploit self-organization stems from their emergence in the course of evolution. The process of trial-and-error in biological evolution can discover beneficial local rules. While some degree of trial-and-error is also involved in the development of technical systems, this process is cut short by rational design. 
It is tempting to exploit self-organization in technical systems as the biological examples show that self-organizing systems are typically highly resilient. However, our ability to rationally design self-organizing systems is limited by our ability to foresee the macroscopic behaviour to which a given set of local interactions leads. Therefore, self-organization is presently not widely exploited in the functioning of technical systems, and if self-organization takes place in these systems the effect is often disruptive. By advancing our ability to foresee the macroscopic results of local interactions, research in complexity may thus enhance our ability to engineer highly robust technical systems. 

Current techniques for inferring the microscopic from the macroscopic include the field of inverse statistical mechanics which uses the language of statistical mechanics to study the emergent behaviour of systems of interacting agents \cite{bialek2012statistical}.
Here we address this challenge from a networks perspective.

A major tool in complex systems research is network modelling \cite{albert2002statistical,newman2003structure,boccaletti2006complex}. 
Depicting a complex system as a network, a set of discrete nodes connected by discrete links, simplifies the constituents of the system but retains the complexity that is inherent to their pattern of interactions. Such models are therefore geared towards analysing the emergence of macroscopic structure from these interactions. 

A macroscopic property that has received particular attention is the degree distribution, the probability distribution of the number of links attached to a randomly drawn node. A challenge is thus to determine to what degree distribution a certain set of local rules leads, or conversely, to create a set of local rules that results in a given degree distribution.
Early works addressed this challenge for particular distributions. For instance seminal papers \cite{scalefree,Price} and a more detailed subsequent analysis \cite{newman2001clustering} showed that linear preferential attachment (see below) leads to power-law degree distributions.
More recently, progress has been made by a class of methods called heterogeneous moment closure approximations \cite{momentrev,vazquez,pugliese,marceau,gleeson}, which capture the time evolution of the numbers of certain classes of motif in the system by an infinite system of ordinary differential equations (ODEs).
Further, we have shown \cite{TripleJump} that the infinite-dimensional ODE systems from heterogeneous approximations can be transformed into a low-dimensional system of PDEs. 

In this paper we show how our previously proposed method \cite{TripleJump} can be used to design sets of local rules that result in a dynamical network that self-organizes to a given target degree distribution.
The proposed method is widely applicable and can be extended to cover other network measures beyond the degree distribution.

\section{Method}\label{sec:method}
We address the following challenge: given a set of permissible dynamical processes and a target degree distribution, we seek to determine the rates of processes that drive the system to the target distribution. The proposed method can be broken into steps as follows:
\begin{enumerate}
\item Describe the evolution of the network using a heterogeneous approximation. This leads to an infinite system of ODEs that describe the temporal evolution of the elements of the degree distribution $p_k$. 
\item Transform the infinite system of ODEs obtained from the heterogeneous approximation into a first-order PDE for the generating function $G(x)=\sum_k p_k x^k$.
\item Transform the desired steady-state degree distribution into its generating function form and substitute into the PDE.
\item Use the resulting expression to determine whether the degree distribution is possible and, if so, obtain the relation between rates that must hold.
\end{enumerate}

The advantage of using the generating function PDE is that it gives one equation, rather than an infinite set, that the rates must satisfy. From this it is possible to straightforwardly extract conditions for the individual processes.

The combination of rates required to produce a particular distribution will typically not be unique.
For degree distributions where the generating function derivatives can be written in terms of the generating function the resulting equation can often be simplified and a set of algebraic conditions can be extracted to solve for the rates. Where no such simplification is possible one can impose further constraints. This reduces the space of possible solutions but makes the set of algebraic conditions derived from the generating function equation more manageable.

Below we compare the target degree distributions to the results from agent-based simulations, which uses the rates derived from the generating function equations. We simulate the network models using an event-driven Gillespie algorithm \cite{gillespie}, the parameters for the particular simulations can be found in the figure captions.

\section{Self-organisation with fixed process rates}\label{sec:constant}
We begin by focusing on the self-organization of networks through processes for which the rate per node or per link (depending on the process) is constant.
Considering only a finite set of such processes places constraints on the degree distributions that can be evolved.
In this setting the proposed method provides a test that determines whether a desired degree distribution can be created by a given set of processes or not.
If the distribution can be created then the method reveals the relative rates of processes that lead to the desired degree distribution. 

We illustrate this procedure in four examples: the Poissonian degree distribution, which we mainly use as an illustrative example, the scale-free, negative binomial and geometric distributions.

The Poisson distribution \cite{ErdosReyni} and scale-free distribution \cite{scalefree} are often used in the modelling of complex systems. The Poisson distribution is used for its mathematical simplicity, while scale-free distributions is found in many real-world networks. The negative-binomial distribution \cite{JohnsonBook} interpolates between different shapes of distributions found in nature, depending on parameter vales. Finally, the geometric distribution is an important special case of the negative-binomial distribution for a specific choice of parameter. We discuss the different degree distributions in more detail below.

We begin by considering a network of discrete nodes connected by unweighted, undirected links (labelled $i-j$, for a link between nodes $i$ and $j$). In this example we assume that we have control over up to 8 processes (chosen relatively arbitrarily, based on previous papers \cite{albert2002statistical,scalefree,SIS,Holme,swarming, demirel2012}).
We define the processes as follows:
\begin{itemize}
\item \emph{Random rewiring}. A link $i-j$ is selected at random, i.e.\ with uniform distribution, and broken. One of the two formerly connected nodes $a \in \{i,j\}$ is chosen randomly with equal probability, and a new link created between $a$ and a target node $b$, where $b$ is chosen randomly from all the nodes in the network that are not currently a neighbour of $a$. The rate (per link) at which random rewiring occurs is $w_{\rm r}$.
\item \emph{Preferential rewiring}. A randomly selected link $i-j$ is broken and one of the two formerly connected nodes $a \in \{i,j\}$ is chosen randomly with equal probability.
A new link is created between the chosen node and a target node $b$, not currently connected to $a$. For the target node $b$ we preferentially select nodes of high degree, such that the probability of a node being chosen increases proportional to their degree.
The rate (per link) at which preferential rewiring occurs is $w_{\rm p}$.
\item \emph{Deletion of links}. A randomly selected link $i-j$ is chosen from the network and deleted. The rate (per link) for the removal of links is $l_{\rm d}$.
\item \emph{Random addition of links}. Two unconnected nodes $i$ and $j$ are picked randomly from the network and a link $i-j$ is formed between them.
The rate (per node) at which random addition of links occurs is $l_{\rm r}$.
\item \emph{Preferential addition of links}. Two unconnected nodes $i$ and $j$ are chosen from the network and a link $i-j$ is formed between them. Both nodes are chosen preferentially, with the probability proportional to the degree of the node.
The rate (per node) at which preferential addition of links occurs is $l_{\rm{p}}$.
\item \emph{Deletion of nodes}. A node is selected at random from the network and deleted, together with all its links.
The rate (per node) for the removal of nodes is $n_{\rm{d}}$.
\item \emph{Random addition of nodes}. A node of (fixed) degree $m$ is added to the network.
The incoming node forms links to $m$ existing nodes in the network, which are chosen at random.
The rate (per node) at which random addition of nodes occurs is $n_{\rm{r}}$. 
\item \emph{Addition of nodes by preferential attachment}. A node of (fixed) degree $m$ is added to the network.
The incoming node forms links to $m$ existing nodes in the network which are chosen preferentially, with probability proportional to their degree.
Hence nodes of higher degree are more likely to form links with the incoming node than nodes of lower degree.
The rate (per node) at which preferential addition of nodes occurs is $n_{\rm{p}}$.
\end{itemize}
\begin{table*}[!t]
	\caption{Target Degree Distributions Produced Using Fixed Process Rates}
	\label{table:table1}
	\centering
$	\begin{array}{lccl}
		\toprule
		\multicolumn{1}{c}{\text{Target distribution}} 
		& p^\star_k
		& G^\star(x)
		& \multicolumn{1}{c}{ \text{Rates}}
		\\
		\midrule
		\text{Poisson} 
		& \dfrac{\mathrm{e}^{-\kay}\kay^k}{k!}
		& \mathrm{e}^{\kay\left(x-1 \right)}
		& 
		\begin{aligned}\kay &= \dfrac{2 l_{\rm{r}}}{l_{\rm{d}}} \quad w_{\rm{r}} = c
		\\
		w_{\rm{p}} &= l_{\rm{p}} = n_{\rm{r}} = n_{\rm{p}} = n_{\rm{d}} = 0 
		\end{aligned}\\
		\midrule
		\text{Power-law}
		& \begin{array}{cr}
    	0 & \text{if } k < m
    	\\
    	\dfrac{2m(m+1)}{k(k+1)(k+2)}  & \text{if } k \geq m
  		\end{array} 
  		& \displaystyle \sum_{k \geq m} \dfrac{2m(m+1)}{k(k+1)(k+2)}x^k
  		& \begin{aligned}n_{\rm{p}} &= c
  		\\
  		l_{\rm{d}} &= l_{\rm{p}} = l_{\rm{r}} = 0
  		\\
  		w_{\rm{p}} &= w_{\rm{r}} = n_{\rm{r}} = n_{\rm{d}} = 0
  		\end{aligned}
  		 \\
		\midrule
		\text{Negative-binomial} & \displaystyle \binom{k + r - 1}{k}p^k(1-p)^r 
		& \displaystyle \left(\dfrac{1 - p}{1 - p x}\right)^r 
		& \begin{aligned}
		p &= \dfrac{\kay w_{\rm p} + 2 l_{\rm p}}{\kay( w_{\rm r} + w_{\rm p} + l_{\rm d} )}
		\\
		r &= \dfrac{\kay(\kay w_{\rm{r}} + 2 l_{\rm{r}})}{\kay w_{\rm{p}} + 2 l_{\rm{p}}}
		\\
		n_{\rm{r}} &= n_{\rm{p}} = n_{\rm{d}} = 0
		\end{aligned} 
		\\
		\midrule
		\text{Geometric} & p(1-p)^k & \dfrac{p}{1-(1-p)x} & 
		\begin{aligned}
		p &= \dfrac{\kay (w_{\rm r} + l_{\rm d} ) - 2 l_{\rm p} }{\kay (w_{\rm r} + w_{\rm p} + l_{\rm d})}\\
		&\text{(subject to the condition)}\\
		0 &= \kay^2 w_{\rm r} + \kay (2 l_{\rm r} - w_{\rm p}) - 2 l_{\rm p}\\
		n_{\rm{r}} &= n_{\rm{p}} = n_{\rm{d}} = 0
		\end{aligned} \\
		\bottomrule
	\end{array}$
\end{table*}
Our goal is to determine rates for the different processes, such that the network degree distribution $p_k$ approaches a target $p_k^*$. We start by capturing the effect of processes 
in a mathematical models. For the processes considered here it is known that the heterogeneous mean field approximation \cite{pastor-satorras} captures the dynamics with good accuracy.  
Using this approximation one derives evolution equation for expectation values of the degree distribution $p_k$ in the limit of large network size $N \rightarrow \infty $. We thereby obtain the following infinite system of ODEs:
\begin{subequations}\label{eq:ExampleEq}
	\renewcommand{\theequation}{\theparentequation \roman{equation}}
	\begin{align}
		\MoveEqLeft\dfrac{{\rm d}p_k}{{\rm d} t} =
		w_{\rm{r}}\left[(k+1)p_{k+1} - kp_k \right.
		\nonumber\\  &
		 \left. \qquad + \left(\textstyle \sum_{k'} k' p_{k'}\right) (p_{k-1} - p_k ) \right]
		\label{eq:ExampleEq_wr}\\ &
		+ w_{\rm{p}} \left[\left((k+1)p_{k+1} - k p_k\right) \right.
 		\nonumber \\ & 
 		\left. \qquad+ \left((k-1)p_{k-1} - k p_k \right) \right] \label{eq:ExampleEq_wp}
		\\ &
		+ l_{\rm{d}} \left[(k+1)p_{k+1} - kp_k \right] \label{eq:ExampleEq_ld}
		\\ &
		+ 2 l_{\rm{r}} \left[p_{k-1} - p_k \right] \label{eq:ExampleEq_lr}
		\\ &
		+ 2 l_{\rm{p}} \left[\left(1/\textstyle\sum_{k'} k' p_{k'}\right)\left((k-1)p_{k-1} - k p_k \right)\right] \label{eq:ExampleEq_lp}
		\\ &
		+ n_{\rm{d}} \left( \textstyle\sum_{k'} k' p_{k'} \right)
		 \left[(k+1)p_{k+1} - k p_k \right] \label{eq:ExampleEq_nd}
		\\ &
		+ n_{\rm{r}} \left[ m(p_{k-1} - p_k) - p_k + \delta_{m,k}  \right] \label{eq:ExampleEq_nr}
		\\&
 		+ n_{\rm{p}} \left[ \left(m/\textstyle\sum_{k'} k' p_{k'}\right)((k-1)p_{k-1} - k p_k)  \right.
 		\nonumber \\ & 
 		\left. \qquad - p_k + \delta_{m,k}  \right]\label{eq:ExampleEq_np},
	\end{align}
\end{subequations}
where $\lbrace w_{\rm{r}}, w_{\rm{p}}, l_{\rm{d}}, l_{\rm{r}}, l_{\rm{p}}, n_{\rm{d}}, n_{\rm{r}}, n_{\rm{p}} \rbrace$ are the rates that we seek to determine, and $\delta_{m,k}$ is the Kronecker delta.

Each line of \eqref{eq:ExampleEq} corresponds to one of the processes and the different terms correspond to different effects of the process.
For rewiring processes \eqref{eq:ExampleEq_wr} and \eqref{eq:ExampleEq_wp}, the term proportional to $(k+1)p_{k+1} - kp_k$ captures the effect of links being rewired away from the focal node; the first term represents the gain in nodes of degree $k$ because of nodes of degree $k+1$ losing one link, while the second represents the loss of nodes of degree $k$ due to such nodes losing one link.
The remaining terms capture the effect of links being rewired to the focal node. This is dependent on the total number of links in the system $\left(\textstyle\sum_{k'} k' p_{k'}\right)$.
Nodes are rewired randomly in \eqref{eq:ExampleEq_wr} and preferentially in \eqref{eq:ExampleEq_wp} where the rate depends on the degree of the node ($k/\textstyle\sum_{k'} k' p_{k'}$).

Adding links (\eqref{eq:ExampleEq_lr} and \eqref{eq:ExampleEq_lp}) occurs at a rate per node, leading to a factor of two in the process rates while deleting links occurs at a per link rate \eqref{eq:ExampleEq_lp}.

There are two ways in which removing nodes \eqref{eq:ExampleEq_nd} can affect the density of $p_k$. Firstly, a neighbour of the focal node can be removed, captured in the evolution equation by the terms in \eqref{eq:ExampleEq_nd} proportional to $(k+1)p_{k+1} - k p_k$, where the factor ($\textstyle\sum_{k'} k' p_{k'}$) is due to the loss of all of the links belonging to the deleted node.
Secondly, the focal node can be deleted, resulting in a decrease in nodes proportional to the density $p_k$.
Since the overall number of nodes in the system has decreased we also need to renormalise the degree distribution resulting in a gain term proportional to $p_k$.
The total change due to deletion of nodes is therefore given by $n_{\rm d} [ - p_k + \textstyle\sum_{k'} k' p_{k'}((k+1) p_{k+1} - k p_{k} )+ p_{k} ]$; we can then cancel the term for removal of the focal node with the renormalisation term.

Lastly, nodes can be added to the network.
We could, if desired, have multiple rules that add nodes of different degrees to the system.
In theory we could allow nodes of every degree to be added to the network, each at a different rate.
In this instance we could trivially create any network model.

For sake of clarity here, to keep the number of processes relatively small, we always add nodes of degree $m$. This increases the density of $p_m$ nodes, leading to the Kronecker delta $\delta_{m,k}$ in \eqref{eq:ExampleEq_nr} and \eqref{eq:ExampleEq_np}.
Nodes of degree $k$ are affected by new nodes forming links to nodes of degree $k$ and $k-1$, this happens randomly \eqref{eq:ExampleEq_nr} or preferentially \eqref{eq:ExampleEq_np} where nodes are selected at a rate that is proportional to the degree of the node $\left(k/\textstyle\sum_{k'} k' p_{k'}\right)$.
Since the new node is of degree $m$ there are $m$ chances for this to happen.
Similar to the case where nodes are deleted from the system, there is also a renormalisation term to account for the change in system size. 

The heterogeneous expansion thus results in an infinite system of ODEs, which we transform into a first-order quasilinear PDE by use of generating functions \cite{genfunctions}.
We start by defining the generating function $G(x,t) = \sum_k p_k(t) x^k$. 
The underlying idea of this transformation is to interpret the elements of the degree distribution as coefficients of a Taylor series of a function $G$ in an arbitrary variable $x$.
This transformation is advantageous because it allows us to work with the continuous object $G$ rather than the discrete set $p_k$.
Because the transformation is reversible (by a Taylor expansion of $G$) no information is lost in the transformation.
Thus investigating the time evolution of $G$ reveals the same information as investigating the time evolution of $p_k$.

To study the time dependence of $G$ we multiply \eqref{eq:ExampleEq} by $x^k$ and sum over $k\geq0$ yielding a first-order PDE for $G(x,t)$
\begin{equation}\label{eq:GPDE}
	\begin{split}
		\MoveEqLeft G_t
		= \left(x - 1 \right) \left[ x \left( w_{\rm p} +  \dfrac{2 l_{\rm p}}{G_x(1,t)}+ \frac{n_{\rm p} m}{G_x(1,t)}\right)
		\right.
		\\ &
		\qquad \left.-  w_{\rm r} - w_{\rm p} - l_{\rm d} - n_{\rm d} G_x(1,t)  \right] G_x 
		\\&
		+ \left[\left(x - 1 \right) \left(w_{\rm r}  G_x(1,t) + 2 l_{\rm r} + n_{\rm r} m\right) \right.
		\\ &
	 	\qquad - \left. n_{\rm r} - n_{\rm p} \right] G
	 	+ \left(n_{\rm r} + n_{\rm p} \right)x^m
	 \end{split}
\end{equation}
where $G_t={\partial G}/{\partial t}$, etc.

To arrive at this equation we broke the right hand side summation into individual sums and then shifted the summation index to turn all instances of $p_{k+1}$ and $p_{k-1}$ into $p_k$. Factors of $x$ can be pulled into or out of the sums as necessary, while factors of $k$ are eliminated using the fact that $\sum k p_k x^{k-1} = \partial_x\sum p_k x^k = G_x$ \cite{genfunctions} leading to the appearance of the spatial derivative in \eqref{eq:GPDE}. Finally we used $G_x(1,t) = \sum_k k p_k(t)$ to eliminate the sums that appear in \eqref{eq:ExampleEq}.

In the present paper we do not attempt to solve this PDE nor to prove existence, uniqueness or stability of solutions but only seek to determine under which conditions it admits a desired solution. Indeed, as we show in Section~\ref{sec:non_constant}, solutions need not be unique or stable.

Given a target degree distribution $p_k^\star$ we can compute the corresponding target generating function 
\begin{equation*}
G^\star(x) = \sum_k p_k^\star x^k.
\end{equation*}
Substituting $G=G^\star(x)$ into \eqref{eq:GPDE} we obtain an algebraic condition that must be met in order for the system to permit the desired degree distribution as a stationary solution. 
\begin{figure*}[!t]
\centering
    \includegraphics[width=0.8\textwidth]{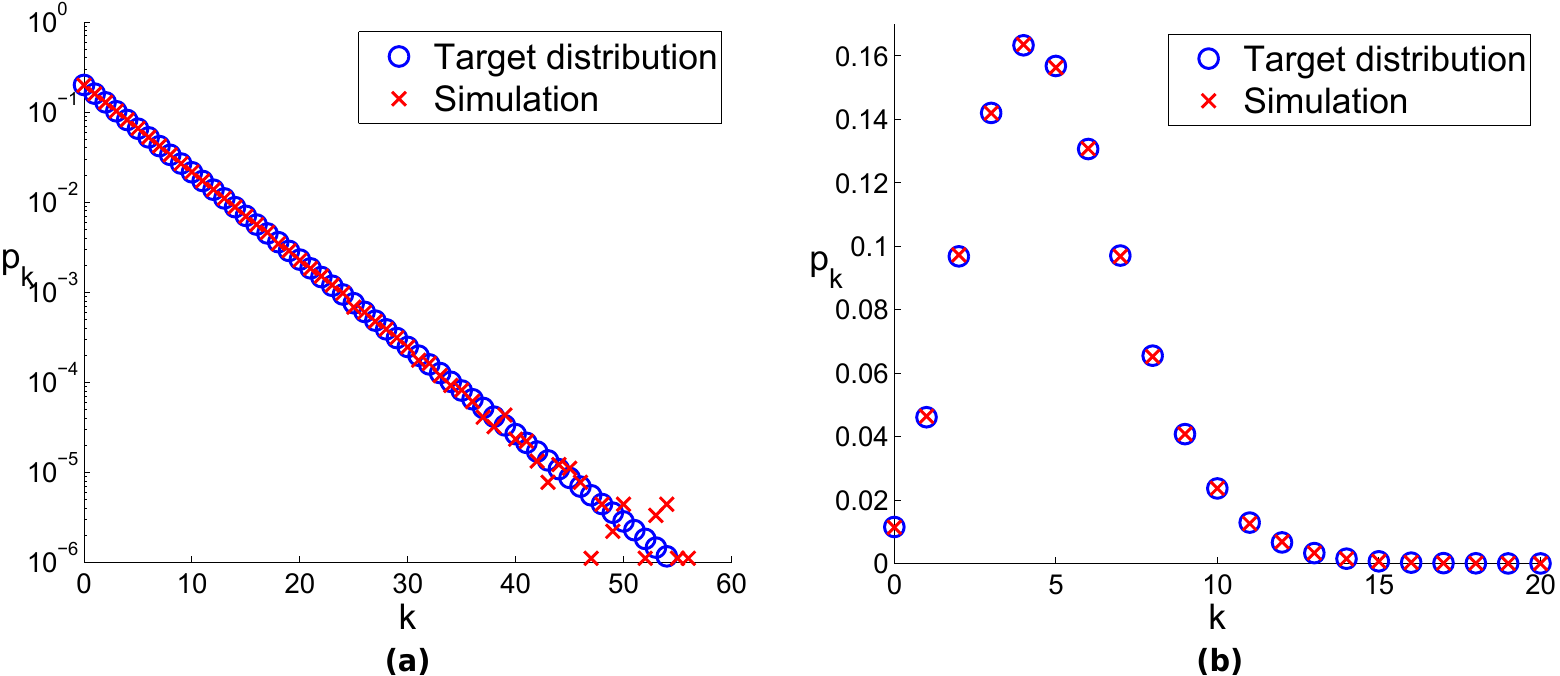}
    \caption{Self-organising networks with local rules achieve target degree distributions. Shown are the target degree distributions (circles) and the self-organised degree distributions in agent-based simulations (crosses). Simulations for a network of size $N = 10^4$ are averaged over 90 runs beginning from three different network configurations (Erd\"{o}s-R\'{e}nyi (ER) network, Barab{\'a}si-Albert (BA) network, and a degree regular (DR) network), with different initial mean degrees ($\kay = 2$, $\kay  = 6$ and $\kay = 8$). (a) The target distribution is long-tailed with $r = 1$,$p = 0.8$ and $\kay = 4$. Hence preferential processes dominate and the corresponding process rates for the simulation are $l_{\rm r} = 0.01$, $l_{\rm p} = 0.04$ and $l_{\rm d} = 0.025$ with all other rates zero (b) The target distribution is more Poissonian with $r = 20$, $p=0.2$ and $\kay = 5$. Hence random processes dominate and the corresponding process rates for the simulation are $l_{\rm r} = 0.04$, $l_{\rm p} = 0.01$ and $l_{\rm d} = 0.02$ with all other rates zero. }
    \label{fig:Rewiring}
\end{figure*}

\subsection{Poisson distribution}
For a simple demonstration we first consider the Poisson distribution $p^\star_k = \exp(-\kay)\kay^k/k!$ \cite{ErdosReyni} as our target distribution, where $\kay$ is the target distribution mean degree. Since the Poisson degree distribution is the degree distribution of a completely random graph, one can guess that this distribution can be created by random rewiring of links or by random addition and deletion of links. To show this using the proposed method we compute the target generating function
\begin{equation}\label{eq:poisson}
G^\star(x)= \e^{- \kay} \sum_k  \dfrac{\kay^k x^k}{k!} = \mathrm{e}^{\kay\left(x-1 \right)}.
\end{equation}
Substituting \eqref{eq:poisson} into \eqref{eq:GPDE} yields
\begin{equation}\label{eq:PoissonPDE}
	\begin{split}
		\MoveEqLeft 0 = \left(x - 1 \right)
		\left[x \left(w_{\rm p} +  \dfrac{2 l_{\rm p}}{\kay}+ \frac{n_{\rm p} m}{\kay}\right)\right.
		\\ &
		\left. \qquad - w_{\rm r} - w_{\rm p} - l_{\rm d}
		 - n_{\rm d} \kay  \right] \kay \e^{\kay\left(x-1 \right)}
		\\ &
		+\left[\left(x - 1 \right) \left(w_{\rm r} \kay + 2 l_{\rm r} + n_{\rm r} m\right) \right. 
		\\ &
		\left. \qquad - n_{\rm r} - n_{\rm p} \right] 
		\e^{\kay \left(x-1 \right)} 
	 	+ \left(n_{\rm r} + n_{\rm p} \right)x^m.
	 	\end{split}
\end{equation}
which must hold for all $x\in\mathbb{R}$. Thus the coefficients of the linearly independent functions $x^2\exp(-\kay(x-1))$, $x\exp(-\kay(x-1))$, $\exp(-\kay(x-1))$, and $x^m$ must all be zero. In particular, then, since the coefficient of $x^m$ must be zero and the rates must be non-negative, we have that $n_{\rm r} = n_{\rm p} = 0$. This implies that there can be no addition of nodes to the network, and hence the rate for removal of nodes must also be zero ($n_{\rm d}=0$) to prevent an absorbing state of an empty network. Under these conditions,  \eqref{eq:PoissonPDE} simplifies to
\begin{equation}\label{eq:PoissonCond}
	\begin{split}
		\MoveEqLeft 0 =
		\kay \left[x \left(w_{\rm p} +  \dfrac{2 l_{\rm p}}{\kay} \right) - w_{\rm r} - w_{\rm p} - l_{\rm d}
		  \right] 
		\\ &
		+\left[\left(w_{\rm r} \kay + 2 l_{\rm r} \right) \right].
	 	\end{split}
\end{equation}
which gives two equations for the remaining rates (as above, the coefficients of the linearly independent functions of $x$ must be zero). These yield  $w_{\rm p} = l_{\rm p} = 0$, $\kay = 2 l_{\rm r}/l_{\rm d} $, and $w_{\rm r} = c$, any constant.
Since the number of links and nodes remains constant for rewiring, the random rewiring rate does not affect the mean degree of the network.
Hence $G^\star(x) = \exp \left[ 2 l_{\rm r}/l_{\rm d} (x-1) \right]$, and the mean degree is the ratio of the rates governing random link addition and link removal.

As expected, the results show that it is possible to design a network with a steady state Poisson distribution with any desired mean degree by choosing rates for random link addition and random link deletion, with a specific quotient.
If $l_{\rm r} = l_{\rm d} = 0$ and the only process acting on the network is random rewiring, then the mean degree remains the same as the initial mean degree of the network, and so $G^\star(x) = \exp\left[ \langle k \rangle(x-1) \right]$ where $\kay$ is the initial mean degree.

Sets of rates that let the network self-organize to other degree distributions can be identified analogously.
We cannot expect to be able to create an arbitrary degree distribution from a finite set of processes running at constant rates.
However, already the set of eight processes considered so far allows us to design networks that self-organise to several common statistical distributions.
We present an overview of some examples in Table~\ref{table:table1} and discuss them briefly below. At the end of the section we also provide an example of a distribution that cannot be achieved with the current set of rules.

\subsection{Power-law distribution}
It is well known that power-law degree distributions with exponent $\gamma=3$ emerge from a process of preferential attachment \cite{scalefree}. Repeating the procedure above with the same set of processes, for such a desired power-law degree distribution, reveals that a network subject to these processes running at constant rates, can only approach a power law degree distribution when addition of nodes by preferential attachment is the only process with non-zero rate. 

\subsection{Negative-binomial distribution}
The negative-binomial degree distribution \cite{JohnsonBook} has two free parameters $p$ and $r$.
When $r=1$ we have a geometric distribution, and when $r\rightarrow \infty$ we recover a Poisson distribution. Applying the proposed method reveals the dependence of the parameters $p$ and $r$ on the rates of processes, shown in Table~\ref{table:table1}, and shows that the distribution is possible whenever there is no addition or deletion of nodes ($n_{\rm d}=n_{\rm r}=n_{\rm p}=0$).
In this case we have five free parameters to meet the two conditions that arise from the method, in order to obtain a network with desired values of $p$ and $r$. Furthermore, $\kay=G_x(1)$, and so we can determine $\kay$ in terms of $p$ and $r$, and hence the process rates. Substituting this relationship into the results for $p$ and $r$ from Table~\ref{table:table1} yields
\begin{equation}\label{eq:NBRelations1}
	\begin{split}
	p&=\dfrac { \left( l_{d}+w_{p} \right) l_{p}+l_{r}w_{p}}{ \left( l_{r}+l_{p} \right)  \left( w_{r}+w_{p}+l_{d} \right) },
\\
	r&=\dfrac { 2(l_r + l_p)(l_p w_r + l_r(w_r + l_d))}{l_d(l_r w_p + l_p(l_d + w_p))},
\\
	\kay &= \dfrac{2(l_r + l_p)}{l_d}.
	\end{split}
\end{equation}
Alternatively, in the case $l_d = l_r = l_p = 0$, where links are neither deleted nor added, $\kay$ is equal to the initial mean degree of the network, and hence an additional free parameter, resulting in 
\begin{align}\label{eq:NBRelations2}
p &= w_p/(w_p+w_r), & r &= \kay w_r/w_p.
\end{align}
It is therefore possible to produce a specific steady state with a desired $p$, $r$ (and possibly $\kay$) by choosing rates to satisfy either \eqref{eq:NBRelations1} or \eqref{eq:NBRelations2}. For purposes of illustration, we choose parameter values that typify the different classes of distribution exhibited by the negative binomial.
Fig.~\ref{fig:Rewiring} compares the results of agent based simulations with the desired target distributions. In Fig.~\ref{fig:Rewiring}(a) we have the long-tailed distribution, while Fig.~\ref{fig:Rewiring}(b) shows the Poisson-like distribution. The simulation results are in good agreement with the target distribution; the discrepancy at high degree in Fig.~\ref{fig:Rewiring}(a) is due to the infrequency of nodes with high degree. This would approach the desired target as the size of the simulation increases.

The final example in Table~\ref{table:table1} is the geometric distribution \cite{JohnsonBook}, which has one free parameter $p$. This is a special case of the negative binomial distribution where $r = 1$.

We have shown that it is possible to produce a number of different degree distributions using the processes of random and preferential rewiring, random and preferential link and node addition, and link and node removal. 
Clearly there are also many distributions that cannot be obtained with the rules considered so far, where applying the proposed method yields conditions that do not admit any solution.

\subsection{A counter example}
As a final (counter) example we thus consider a degree distribution of the form $p_k^{\star} = \exp(-1)(k+1)/2 k !$, which has corresponding generating function
\begin{equation} \label{eq:counterGF}
G^{\star} = \frac{1}{2}(1 + x) {\rm e}^{x-1}.
\end{equation}
We proceed as before and substitute the target generating function $G^*$ into \eqref{eq:GPDE}, and inspect the coefficients of the linearly independent functions. We find that for the coefficients to be equal to zero we must have a network where nodes are neither added nor removed.
We are left with a simplified equation to solve for the remaining rates
\begin{equation}\label{eq:counterPDE}
	\begin{split}
\MoveEqLeft 0 =  \left[ x \left( w_{\rm p} +  \dfrac{2 l_{\rm p}}{\kay}\right)
- w_{\rm r} - w_{\rm p} - l_{\rm d} \right]
(2 + x)
\\
 & + \left( w_{\rm r} \kay + 2 l_{\rm r} \right)
 (1 + x).
 \end{split}
\end{equation}
In order to satisfy this equation for all $x \in \mathbb{R}$,  all rates must be zero. To see this, note that the coefficient of the $x^2$ term implies that the preferential rates $w_{\rm p}$ and $l_{\rm p}$ must both be zero. The resulting equation $\left ( w_{\rm r} +  l_{\rm d} \right) (2 + x) =\left( w_{\rm r} \kay + 2 l_{\rm r} \right)(1 + x) $, leaves two conditions that cannot both be satisfied
\begin{align*}
 \kay w_{\rm r} + 2 l_{\rm r} &= w_{\rm r} + l_{\rm d} 
 \\
  \kay w_{\rm r} + 2 l_{\rm r} &= 2 (w_{\rm r} +  l_{\rm d}) . 
\end{align*}
The target distribution $G^{\star}$ is therefore not possible under the given set of processes.

In such cases we have two options.
First, we can expand the set of processes by allowing one or more additional processes.
If we continue to allow the processes to only run at constant rates we restrict the types of terms that can appear in the generating function equation.
Processes based on selecting a node will depend on the generating function $G$ or the derivative $G_x$ if selected preferentially. Processes based on selecting a link will depend on the derivative $G_x$. Similarly selecting higher order motifs with result in the inclusion of higher order derivatives into the generating function PDE. For example selecting triplets will result in second derivative terms $G_{xx}$.
We can investigate whether such derivative terms can help in the generating function equation.

If it is not immediately apparent whether the addition of extra terms will help, then we can instead relax the assumption that the processes run at constant rates.
From a node-based perspective, processes that select links or triplets are already running at non-constant rates: the rates depend on the degree $k$ of the nodes when selecting links, and depend on the degree of the nodes like $k(k-1)$ when selecting triplets.

While link-based rules and triplet-based rules lead to processes we can write in terms of the generating function $G$, not all processes will lead to such results. For example, selecting nodes at a rate proportional to $k/(k+1)$, does not have an obvious generating function equivalent. We cannot express the term $\sum_k k p_k /(k + 1) x^k $ in terms of $G$ and its derivatives.
We discuss such non-linear processes and how to identify them in the next section. We also show how such non-linear rates can lead to a  network that evolves towards \eqref{eq:counterGF}.
\section{Networks with degree-dependent rates}\label{sec:non_constant}
\begin{table*}[t]
\caption{Target Degree Distributions Produced Using Degree Dependent Rates}
	\label{table:table2}
	\centering
		$\begin{array}{cccl}
		\toprule
		\text{Target distribution} & p^\star_k & G^\star(x) & \multicolumn{1}{c}{\text{Rates}}
		\\
		\midrule
		\text{Poisson} 
		&\dfrac{\mathrm{e}^{-\kay}\kay^k}{k!}
		& \mathrm{e}^{\kay\left(x-1 \right)}
		& l_k = \dfrac{m_{k-1} k}{\kay}
		\\
		\midrule
		& \dfrac{(k+1)\mathrm{e}^{-a}a^k}{(1+a) k!}
		& \dfrac{1+ax}{1+a}\mathrm{e}^{a( x - 1)}
		& l_k = \dfrac{k^2(m_{k-1} + \bar{T})}{a(k+1)} - \dfrac{k \bar{T}}{\kay}
		\\
		\midrule
		\text{Power-law}
		&
		\begin{array}{cc} 
		c & \text{if } k = 0 \\ 
		\dfrac{(1-c)k^{-\alpha}}{\zeta(\alpha)} & \text{if } k \geq 1
		\end{array} 
		& c + \dfrac{(1-c)\mathrm{Li}_{\alpha}(x)}{\zeta(\alpha)} 
		& \begin{aligned}
		l_0 &= 0
		\\
		l_1 &= \dfrac{c\zeta(\alpha)(m_0 + \bar{T})}{1-c} - \dfrac{\bar{T}}{\kay}
		\\
		l_k &= \dfrac{(k-1)^{-\alpha}(m_{k-1} + \bar{T})}{k^{-\alpha}} - \dfrac{k \bar{T}}{\kay}, \; k \geq 2
		\end{aligned}
		\\
		\midrule
		\text{Bimodal}
		& \dfrac{\mathrm{e}^{-a} a^k + \mathrm{e}^{ -b } b^k}{ 2 k !} 
		& \dfrac{\mathrm{e}^{a(x-1)} + \mathrm{e}^{b(x-1)}}{2}
		& 
		\begin{aligned}
		l_k &= \dfrac{k (m_{k-1} + \bar{T}) \left( \mathrm{e}^{-a} a^{k-1} + \mathrm{e}^{-b} b^{k-1} \right)}{ \left(\mathrm{e}^{-a} a^k + \mathrm{e}^{-b} b^k \right)}
		 - \dfrac{k\bar{T}}{\kay}
		\end{aligned}  \\
		\bottomrule
	\end{array}$
\end{table*}
Up to this point we have assumed that processes occur at constant rates (per node or per link) that are independent of the respective node's or link's properties. By contrast, in many systems studied in nature rates depend on node properties, such as the node's degree. Also, in technical applications it is easily conceivable that the nodes are aware of their own degree and take it into account in their behaviour. We therefore consider degree-dependent rates in the context of the method proposed here. 

Allowing degree-dependent rates greatly increases the range of degree distributions that can be obtained with a given number of processes, which enables us to restrict the set of processes considered. For illustration we only consider degree-dependent link creation and deletion.

Two variants of degree-dependent link creation/deletion processes are conceivable: using either non-local or local information. In the first, non-local, variant the task of creating a network with given degree distribution is trivial, as we end up with the configuration model \cite{molloy}. Furthermore, the non-local variant requires non-local knowledge to be available at each node and is hence infeasible in many technical applications. We therefore do not consider the non-local degree-dependent processes here. 

Instead we consider local degree-dependent link creation and deletion processes. In this local variant, the decision to create or delete a link is made by the nodes independently, taking only their own degree into account. If a node decides to delete a link it chooses the link randomly among its existing links. If a node decides to create a link it establishes the link to another node that is randomly selected from the whole population. Thus nodes are also subject to link creation and deletion events by partners, which are not under their control.

The time evolution of the degree distribution $p_k$, when only considering link creation and deletion, is captured by 
\begin{subequations}\label{eq:2ODE}
\renewcommand{\theequation}{\theparentequation \roman{equation}}
	\begin{align}
		\MoveEqLeft \dfrac{{\rm d} p_k}{{\rm d} t}
		= - l_k p_k 
		+ l_{k+1} p_{k+1}\label{eq:DelLinki}
		\\ &
		+ \dfrac{\sum_k p_k l_k}{\sum_k k p_k} 
		\left[ (k+1) p_{k+1} - k p_k \right] \label{eq:DelLinkii}
		\\ &
		- m_k p_k 
		+ m_{k-1} p_{k-1}\label{eq:AddLinki}
		\\ &
		+ \sum_k p_k m_k 
		\left[p_{k-1} - p_k \right]\label{eq:AddLinkii}.
	\end{align}
\end{subequations}
The terms in \eqref{eq:2ODE} describe the change in $p_k$ due to the removal of links at a rate $l_k$ and addition of links at a rate $m_k$. Terms \eqref{eq:DelLinki} and \eqref{eq:AddLinki} are due to the focal node, of degree $k$, having a link deleted/added, while \eqref{eq:DelLinkii} and \eqref{eq:AddLinkii} are due to a neighbour, of any degree, adding or removing a link to the focal node.

We now define three generating functions.
The first is the generating function for the degree distribution $p_k$, $G(x,t) = \sum_k p_k(t) x^k$, while the remaining two represent the degree distribution multiplied by the link removal rate, $S(x,t) = \sum_k l_k p_k(t) x^k$ and the link addition rate, $T(x,t) = \sum_k m_k p_k(t) x^k$.
The need to define two new generating functions stems from the non-constant process rates; when these rates are multiplied by the degree distribution the result will not in general be a multiple of the generating function $G$. The form of the new generating functions is chosen to make the transformation of \eqref{eq:2ODE} to a generating function PDE straightforward. Multiplying (\ref{eq:2ODE}) by $x^k$ and summing over $k \geq 0$ gives the first-order PDE
\begin{multline}\label{eq:NonConstPDE}
	G_t 
	= S\left( \dfrac{1}{x} - 1 \right)
	+ T ( x - 1) 
	\\
	+ \dfrac{S(1)}{G_x(1)} \left( 1 - x \right)G_x
	+ T(1)\left( x - 1 \right)G.
\end{multline}
In the steady state this simplifies to
\begin{equation}\label{eq:2SS}
S = x\left( T + \bar{T}G  - \dfrac{\bar{S} G_x}{\kay} \right),
\end{equation}
where $\bar{S} = S(1)$ is the total rate of link addition events per node, $\bar{T} = T(1)$ is the total rate of link deletion events, and $\kay$ is the mean degree as above.

Since we do not consider node additions or deletions, the degree distribution can only be stationary if the total link addition and deletion rates are identical. We can verify this by evaluating \eqref{eq:NonConstPDE} at $x=1$. Since $G_x(1)=\kay$, $T(1)=\bar{T}$ and $S(1)= \bar{S}$ we find $\bar{T}=\bar{S}$ as expected.

As before, we have a great deal of freedom when specifying the link rates.
Typically one first chooses $m_k$ which in turn determines
 $l_k$, where one must be careful to check that the particular choice of $m_k$ does not result in negative values for $l_k$.

For simplicity, we again consider which combinations of processes can lead to the Poisson distribution, which has desired degree distribution $p^\star_k = \exp(-\kay) \kay^{k}/k!$, and hence $G^\star(x) = \exp \left[\kay(x-1)\right]$. Substituting $G=G^\star(x)$ into \eqref{eq:2SS} yields
\begin{equation}\label{eq:2Poisson}
	S = x \left[T + \left(\bar{T}-\bar{S}\right)G^\star\right].
\end{equation}
Since $\bar{S} = \bar{T}$, we can cancel the two terms in \eqref{eq:2Poisson} and are left with the relationship $S = xT$. By comparing coefficients of $x^k$ we find the condition $l_k=m_{k-1} (p_{k-1}/p_k)$ and hence $l_k  = k m_{k-1}/\kay$.

We can use this relationship to reproduce a result from the previous section. If links are added independently of degree, e.g. $m_k=1$, the required loss rate is $l_k=k/\kay$. So links are lost proportionally to a node's degree, which means a fixed-rate link loss per link, which leads to the same system identified above. 

This solution is not unique. For example, if we allow links to be added at a rate proportional to degree, so $m_k = k$, then $l_k = k (k-1)/\kay$, such that loss is proportional to the number of distinct pairs of links connecting to a node.

The above analysis can be repeated with other distributions.
Some examples are listed in Table~\ref{table:table2} (where  $\zeta(\alpha)$ is the zeta function and $\mathrm{Li}_{\alpha}(x)$ is the polylogarithm of $x$).
Once we have a relationship between $l_k$ and $m_k$, as given in Table~\ref{table:table2}, we can choose values for $m_k$ (or $l_k$) and hence calculate $\bar{T}$ in order to find the corresponding $l_k$ (or $m_k$).

The second distribution in Table~\ref{table:table2} is a more general version of \eqref{eq:counterGF} from Section~\ref{sec:constant} where we had $a = 1$.
While we were unable to produce the distribution with constant process rates we see that it is now possible to produce the target distribution by adding and deleting links at rates that depend on the degree of the node.

\begin{figure}
\centering
    \includegraphics[width=0.35\textwidth]{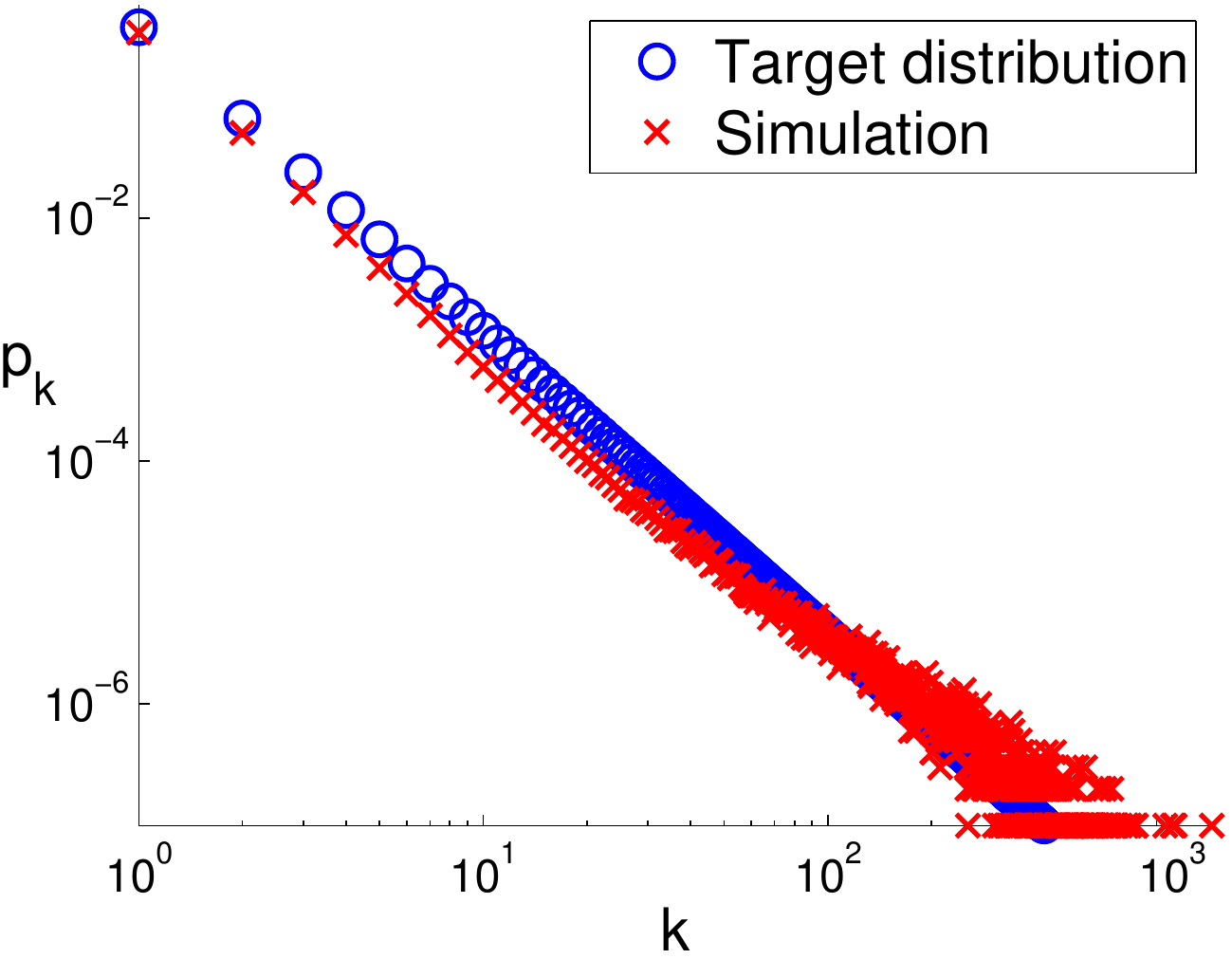}
    \caption{Self-organizing network with degree-dependent process rates.
Using only link creation and link removal, functional-forms for the degree-dependence of rates were designed such that the network approaches a power-law degree distribution with exponent -2.5. Agent-based simulations (crosses) show that the designed system approximate the target distributions (circles).
We simulate a system of size $N = 10^5$ averaged over 100 runs from two different initial network configurations (ER-network and DR-network) and initial mean degree ranging from $\kay  = 1$ to $\kay =3$}
    \label{fig:DegKRules}
\end{figure}

Table~\ref{table:table2} also gives the condition for a power law degree distribution with exponent $\alpha$ and given $p_0=c$ to prevent divergence of the distribution at $k=0$.
A comparison between the target distribution, with $\alpha = 2.5$ and $c = 0.5$, and an agent-based simulation is shown in Fig.~\ref{fig:DegKRules}(a).
Using the rules in Table~\ref{table:table2} we simulate the network by adding links to nodes at a rate proportional to their degree, choosing $m_k = 0.05 k$, thus in accordance with the conditions in  Table~\ref{table:table2} delete links at the rates,
\begin{align*}
l_1 &= 0.025(\zeta(\alpha - 1) - 2),
\\
l_k &= 0.05\left(k-1 + \frac{\zeta(\alpha - 1)}{2 \zeta(\alpha)}\right)\left(\frac{k}{k - 1}\right)^{\alpha} - 0.05 k, 
\\ & \qquad  k \geq 2.
\end{align*}
The results from the agent-based simulation produce the power-law shape of the target distribution. We expect the accuracy to increase as the size of the simulation increases.

\begin{figure*}[!t]
\centering
    \includegraphics[width=0.8\textwidth]{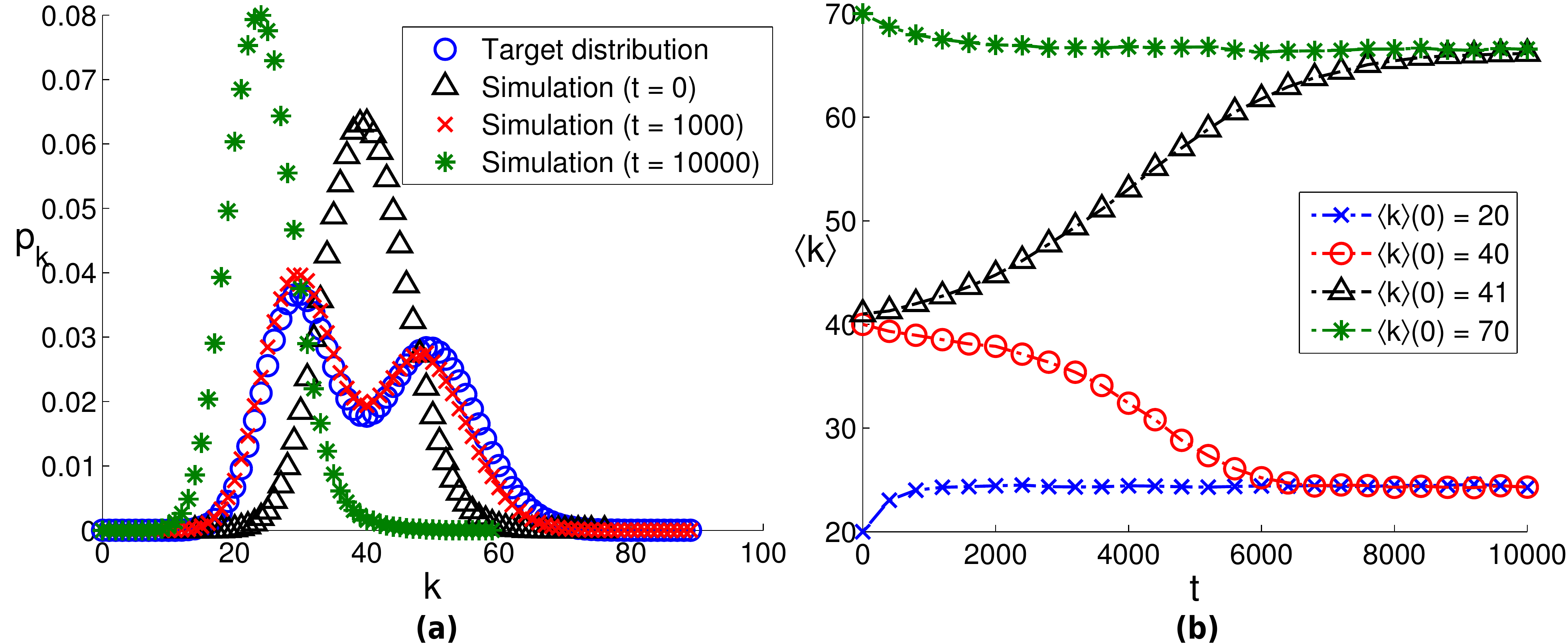}
    \caption{Self-organizing network with degree-dependent process rates. Agent-based simulations (crosses) for a system of size $N = 10^4$ show that the designed system approximates the target bimodal distributions (circles), which is unstable. Starting from the target mean degree ($\kay = 40$) the distribution approached the target distribution (at $t = 1000$) before moving towards a different, unimodal, stable steady-state (a).
The system never approaches the target distribution as $t \rightarrow \infty$;  
(b) shows typical trajectories of the mean degree for different initial mean degree. Networks are initialised as an ER-network and simulations are averaged over 90 runs in (a). }
    \label{fig:bimodal}
\end{figure*}

Table~\ref{table:table2} also contains an example of an unstable distribution.
A comparison between a target bimodal distribution, where $a = 30$ and $b = 50$, and agent-based simulations is shown in Fig.~\ref{fig:bimodal}(a).  
Using the process rates from Table~\ref{table:table2} we simulate the network model by adding links at a constant rate, $m_k = 0.04$ and hence delete links at a rate
\[l_k = \dfrac{0.08 k \left( (0.6)^{k-1} + {\rm e}^ {- 20 } \right) }{50 (0.6^k +{ \rm e}^ { - 20 }) } - \dfrac{0.04 k}{\kay} .\]

When starting with a mean degree that is close to the target distribution mean degree, the agent-based model initially approaches the target distribution before moving towards a different stable steady-state with lower mean degree (Fig.\ref{fig:bimodal}(a)).
Simulations that begin with a mean degree greater than or less than the target distribution never reach the target mean but instead move towards one of two different steady-states, which are both unimodal.
Fig.\ref{fig:bimodal}(b) plots typical trajectories of the mean degree over time from different initial mean degrees.

The above example shows that though we can design a network to self-organise towards a desired steady state there is no guarantee that the target state is a stable solution of the generating function PDE.
Furthermore, the resulting degree-dependent rates do not necessarily lead to a closed form function in terms of the generating function $G$. In this case, analysis of the generating function PDE is virtually impossible. However, as we have shown, once the solution exists one can verify the results through simulation.
We therefore think of the method as it currently stands as a two stage process that can be used to find feasible targets.
The first step uses the generating function PDE to investigate whether a target is viable under a given set of rules.
Any solution that is then considered for implementation in the real world can be tested in simulations, where stability can be examined.
\section{State-change processes}\label{sec:binary}

In applications, the self-organisation of a dynamical network may involve the assignment of functional roles to the nodes. For instance one can imagine a self-organizing sensor network\cite{leonard2004}, where initially identical smart sensors differentiate into two functional states, say primary recorders of data and aggregators, who integrate data from different recorders and transmit results. In this case we may want the system to evolve a communication network where the aggregators are hubs that connect to many recorders and some other aggregators. 

In this section we address the challenge of designing a self-organizing network where both the states of nodes and the state-dependent degree distributions approach predefined targets. 
We proceed as before and define a set of processes acting on the network and state-dependent degree distributions and frequencies of the different states. We then describe the evolution of the network using a heterogeneous active-neighbourhood approximation \cite{marceau,lindquist}, which tracks the evolution of nodes in a specific state and the number of neighbours it has in each state.
When we have a single-state network the active-neighbourhood approximation and the heterogeneous mean-field approximation of the previous sections are equivalent.

The active-neighbourhood approximation results in coupled infinite-dimensional systems of ODEs, which we then convert into coupled PDEs using generating functions.
For a system with $N$ distinct node states we obtain a system of $N$ coupled PDEs. Even for systems with several states this does not pose a fundamental problem as we do not need to solve the PDEs.
By substituting the target degree distributions into the PDE system we find the conditions that the process rates must satisfy to reach the desired target.

For illustration we consider a challenge inspired by the sensor network example.
Our aim is to determine rules that self-organize the network to a state where a given proportion of the nodes become aggregators, state $A$, while the others become recorders, state $B$.
Furthermore we want the aggregators connected among themselves in a network with a Poissonian degree distribution with a desired mean, similarly for the recorder to recorder connections and the aggregator to recorder connections.

We define six dynamical processes acting on the network comprising link-rewiring and state-change processes, with constant rates $w_{\rm p}$ and $p_{\rm p}$ for a process ${\rm p}$ respectively, as described below. A node in state $i\in\{ A,B\}$ can rewire an existing link from a neighbour in state 
$j\in\{A,B\}$ to a node in the other state $\bar{j}$, picked uniformly at random from the network.
There are four such rewiring processes; the rates at which they occur are denoted as $w_{\rm ij-i\bar{j}}$.
The remaining two processes are state-change processes; a node in state $i\in\{ A,B\}$ can switch to the opposite state $\bar{i}$, the rate at which these processes occur are $p_{\rm i-\bar{i}}$.

We define $N_{k,l}$ as the density of nodes in state $N \in \lbrace A,B\rbrace$ with $k$ $A$-neighbours, and $l$ $B$-neighbours. The evolution of the density of $A_{k,l}$ nodes and $B_{k,l}$ nodes under the six processes described above results in two coupled infinite-dimensional systems of ODEs, the equations are given in Appendix~\ref{appendix:A}.

We next introduce the generating functions $G^A = \sum_{k,l}A_{k,l}x^k y^l$ and $G^B = \sum_{k,l}B_{k,l}x^k y^l$, and convert the pair of infinite-dimensional systems of ODEs into two first-order coupled PDEs; the equations are given in Appendix~\ref{appendix:B}. 

We substitute target steady-state degree distributions into the steady-state generating function equations and compare coefficients of linearly independent functions, as before, in order to find the necessary relationships between rates.
\begin{figure*}[!t]
\centering
    \includegraphics[width=0.9 \textwidth]{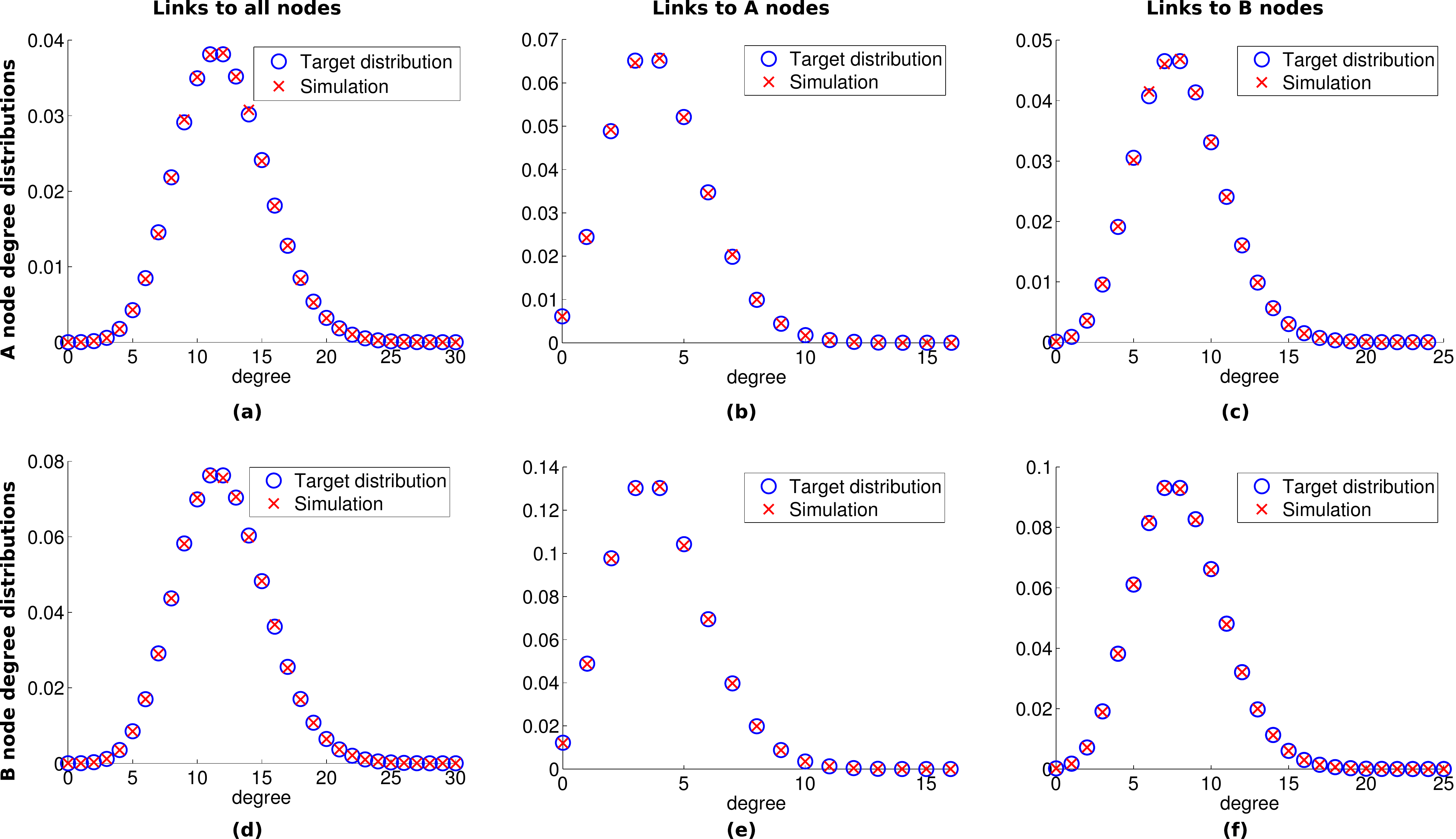}
    \caption{Local rules generate target distributions in a two-state network. Shown are target distributions (circles) compared to agent-based simulations (crosses) designed to self-organize to the target distribution by using the relations \eqref{eq:BinaryRates1}. Rates are as follows: $w_{\rm AB-AA}  = 0.01 , w_{\rm AB-BB} = 0.02, w_{\rm AA-AB} = 0.04, w_{\rm BB-AB} = 0.02, p _{\rm A-B}  = 0.03 , p_{\rm B-A} = 0.015$. Top is the degree distribution of $A$-nodes where (a) is the total degree distribution (b) is the degree distribution to $A$-nodes only (c) is the degree distribution to $B$-nodes only. Bottom is the degree distribution of $B$-nodes where (d) is the total degree distribution (e) is the degree distribution to $A$-nodes only (f) is the degree distribution to $B$-nodes only. We simulated a system of size $N=10^4$ averaged over 90 runs from three initial network configurations (ER-network, BA-network and DR-network) and five initial B-node fractions, $B_0 = \lbrace 0.01, 0.1, 0.5, 0.9, 0.99 \rbrace $.}
    \label{fig:Binary1}
\end{figure*}
In our sensor network example, we thus define two generating functions: one for the aggregators $G^A(x,y) = c_1 \exp [ a (x - 1) + b (y - 1) ]$, and one for the recorders $G^B(x,y) = c_2 \exp [ a (x - 1) + b (y - 1) ]$.

The exponents $a$ and $b$ are common between $G^A$ and $G^B$ for simplicity; we shall relax this constraint below. Here $c_1$ is the proportion of aggregators and $c_2$ is the proportion of recorders, and hence $c_1 + c_2 = 1$ is the total density of sensors. The average number of aggregator to aggregator (or aggregator to recorder) connections per aggregator (or recorder) is $a$, while the average number of recorder to aggregator (or recorder to recorder) connections per aggregator (or recorder) is $b$. The choice of values for $a$, $b$, $c_1$ and $c_2$ is constrained by the condition $c_1 b = c_2 a$, which ensures symmetry; the number of $AB$-links must be equal to the number of $BA$-links; this can be equivalently written as $G^A_y(1,1) = G^B_x(1,1)$.

\begin{figure*}
\centering
    \includegraphics[width=0.9 \textwidth]{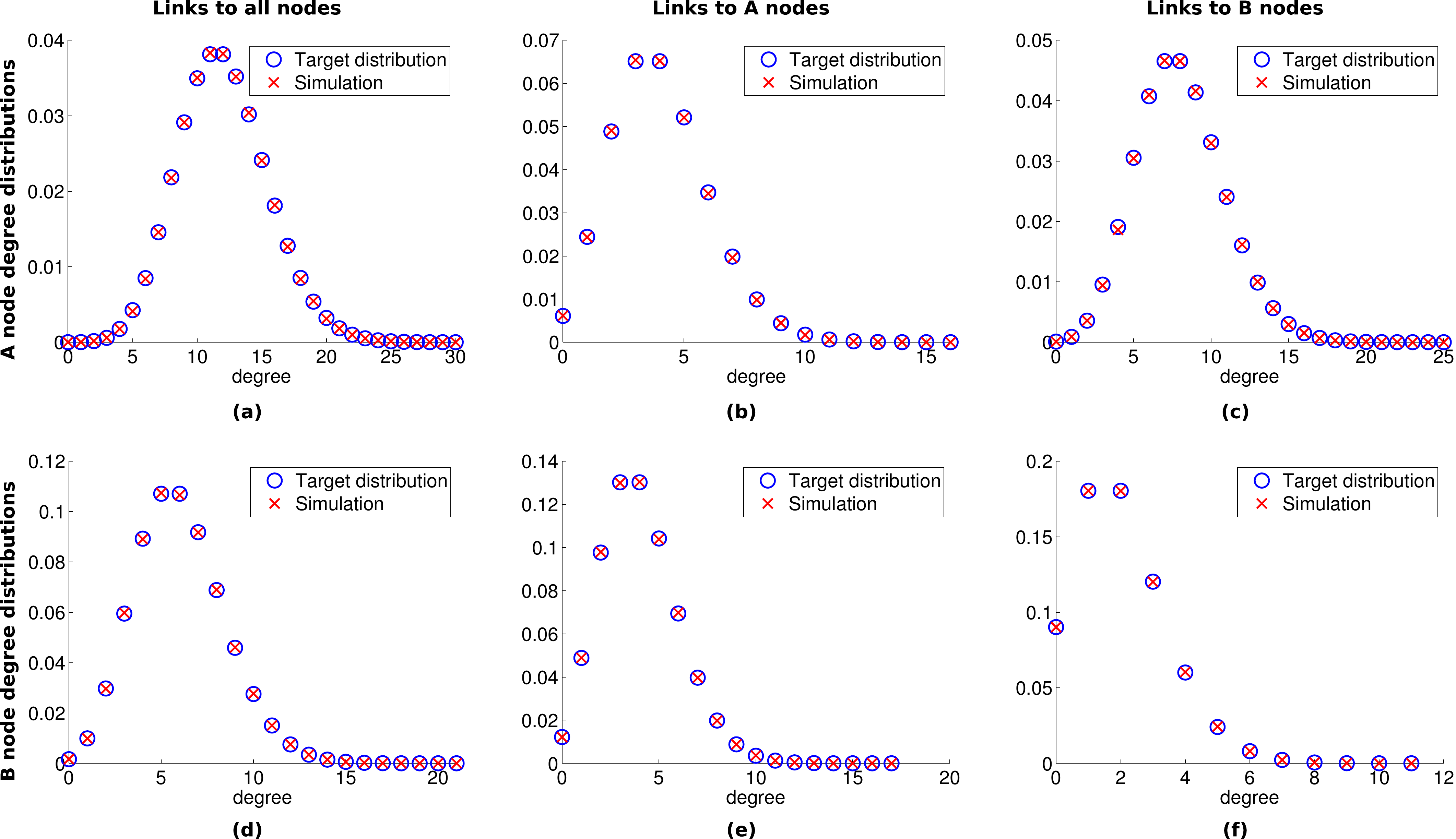}
    \caption{Local rules generate target distributions in a two-state network; general case. Shown are target distributions (circles) compared to agent-based simulations (crosses) designed to self-organize to the target distribution by using the relations \eqref{eq:BinaryRatesWire} and \eqref{eq:BinaryRatesState}. Rates are as follows: $w_{\rm AB-AA}  = 0.01 , w_{\rm AB-BB} = 0.02, w_{\rm AA-AB} = 0.04, w_{\rm BB-AB} = 0.08, \alpha_{k,l} = 0.05, \beta_{k,l} = 0.025 \exp(-6)(4)^l $.  Top is the degree distribution of $A$-nodes where (a) is the total degree distribution (b) is the degree distribution to $A$-nodes only (c) is the degree distribution to $B$-nodes only. Bottom is the degree distribution of $B$-nodes where (d) is the total degree distribution (e) is the degree distribution to $A$-nodes only (f) is the degree distribution to $B$-nodes only. Note differences in vertical scales. We simulated a system of size $N=10^4$ averaged over 90 runs from three initial network configurations ((ER)-network, (BA)-network and (DR)-network) and five initial B-node fractions, $B_0 = \lbrace 0.01, 0.1, 0.5, 0.9, 0.99 \rbrace $.}
    \label{fig:Binary2}
\end{figure*}
Following the proposed method, we substitute $G^A$ and $G^B$ into the steady-state generating function equations in Appendix~\ref{appendix:B}. We are able to cancel the exponential function $\exp[a (x - 1) + b (y - 1)]$ and then compare coefficients of $x$ and $y$.
We find
\begin{equation}
\begin{split}\label{eq:BinaryRates1}
p_{\rm A-B} - \dfrac{b}{a}p_{\rm B-A} & = 0
\\
w_{\rm AB-AA} - \dfrac{a}{b}\dfrac{w_{\rm AA-AB}}{2} & = 0
\\
w_{\rm AB-BB} - \dfrac{b}{a}\dfrac{w_{\rm BB-AB}}{2} & = 0.
\end{split}
\end{equation}
Thus there is a wide range of feasible choices of process rates to satisfy these conditions for any given target distribution, with parameters $a$ and $b$. 
A comparison between target distributions $G^A = \exp[4(x + 2y -3)]/3$ and $G^B = 2\exp[4 (x + 2y - 3)]/3$ and agent-based simulations subject to the relations \eqref{eq:BinaryRates1}, are shown in Fig.~\ref{fig:Binary1}. The simulation results are in good agreement with the target degree distributions.

The aggregators $A$, in our sensor network are less abundant than the data recorders, $B$.
There are many recorders per aggregator and there is low connectivity between aggregators, but high connectivity between recorders.
This is due to our choice of $G^A$ and $G^B$ having equal exponents, hence the connectivity between aggregators and recorders which we wanted to be high is the same as the connectivity between recorders.

It could be advantageous in certain applications for the recorders to be connected with lower mean, potentially leading to the deployment of sensors over a larger area.
We thus define two new target generating functions $G^A(x,y) = c_1 \exp[a_1 (x - 1) + a_2( y - 1)]$ and $G^B(x,y) = c_2 \exp[b_1 (x - 1) + b_2 (y - 1)]$, such that the proportion of aggregators ($c_1$) is less than the proportion of recorders ($c_2$) and the mean of sensors connected of the same type ($a_1$ and $b_2$) is small, while the number of recorders per aggregator is large. Again the parameters are subject to constraints of symmetry and total aggregator and recorder density, which imply $c_1a_2=c_2b_1$ and $c_1+c_2=1$ respectively.

In order to design such a system we must introduce new processes.
As in Section~\ref{sec:non_constant}, we can use the same methodology when we allow for processes that can depend on the degree of the node.
We therefore allow the state-change processes to be degree dependent.
$A$-nodes can change state at rates $\alpha_{k,l}$ and $B$-nodes at rates $\beta_{k,l}$.
As before, we must introduce two new generating functions $S(x,y) = \sum_{k,l}\alpha_{k,l} A_{k,l}x^k y^l$ and $T(x,y) = \sum_{k,l} \beta_{k,l}B_{k,l} x^k y^l$ for the state-change processes.
The steady-state generating function PDEs for a network subject to these processes are given in Appendix~\ref{appendix:C}.

Substituting the target generating functions $G^A$ and $G^B$ into the PDEs gives two equations in six unknowns.
This shows that the system is still under-determined and we have the freedom to impose additional constraints to arrive at a  solution.
Hence here we solve for the rewiring processes and state change processes separately.

For the rewiring equations, we can cancel the generating functions and compare coefficients of $x$ and $y$ to get the following relations between the rewiring rates
\begin{equation}
	\begin{split}\label{eq:BinaryRatesWire}
	w_{\rm AB-AA} -  \dfrac{a_1 }{a_2}\dfrac{w_{\rm AA-AB}}{2}   &= 0,
	 \\
	w_{\rm AB-BB} - \dfrac{c_2 b_2}{c_1 a_2}\dfrac{w_{\rm BB-AB}}{2}  &= 0.
	\end{split}
\end{equation}
Next, solving for the state-chance processes gives the relation between the state change rates $\alpha_{k,l}$ and $\beta_{k,l}$
\begin{equation}\label{eq:BinaryRatesState}
\alpha_{k,l} - \e^{a_1 + a_2 - b_1 - b_2} \dfrac{c_2}{c_1}\left( \dfrac{b_1}{a_1}\right)^k \left(\dfrac{b_2}{a_2}\right)^l \beta_{k,l} = 0.
\end{equation}
Comparisons between target distributions, $G^A = \exp[ 4(x + 2y -3)]/ 3$ and $G^B = 2 \exp[(4x + 2y - 6)]/3$, and agent-based simulations subject to the relations \eqref{eq:BinaryRatesWire} and \eqref{eq:BinaryRatesState} are shown in Fig.~\ref{fig:Binary2}.
Compared to Fig.~\ref{fig:Binary1}, the connectivity between aggregators and between aggregators and recorders remains the same, but there are fewer recorder to recorder connections, as per the design criteria.
The results from the agent-based simulation are in good agreement with the target degree distributions.

\section{Conclusion}
In this paper we proposed a method for the design of rules that let a network self-organize into a target steady-state degree distribution.
This is achieved by first modelling the network using a heterogeneous moment expansion.
The infinite-dimensional system of ODEs from the heterogeneous approximation can then be converted into first order PDEs using generating functions, where the number of PDEs will depend on the number of states in the system. By substituting the target steady-state degree distribution into the generating function PDEs we derive algebraic consistency conditions, from which it is possible to determine which processes on a network result in the target degree distribution.

There are a number of caveats to the method proposed here, which concern the convergence to the desired state, the validity of the approximation and the applicability in the real world.
First, the method proposed here generates a set of rules under which the desired state is stationary.
However, it does not guarantee that this state is locally dynamically stable or globally attractive.
For systems with degree-independent rules the global attractivity should not present a problem as these rules lead to linear systems, which have only a single attractor.
For non-linear degree-dependent processes, as we saw in Section~\ref{sec:non_constant}, multiple attractors can exist, thus global attractivity is hard to guarantee.
However, the example of Pyragas control \cite{pyragas1992}, for instance, shows that methods which only guarantee the existence but not stability of a solution can be useful in practise.
Such methods, including the one here, can be used in the design stage to narrow down the space of possible solutions.
Any solution that is then considered for implementation in the real world will certainly first be tested in simulations, where local and global stability can be examined. 

A second concern is the mathematical validity of the approach.
The present implementation of the method is exact except for the active-neighbourhood approximation.
This approximation is known to provide a highly accurate approximation for stationary states of dynamical networks \cite{marceau}.
The approximation relies on the absence of long-ranged correlation in the network.
Such correlations can arise during transients, which is of little concern, and in certain systems close to bifurcations.
As a general rule, detrimental correlations will be present, first, when the network fragments on a global scale (such as the fragmentation transition in the adaptive voter model \cite{motif,momentrev}), or, second, when processes in the network lead to an over-abundance of certain meso-scale motifs, that far exceeds statistical expectations. 

The logical extension of the method, beyond the implementation presented here will be 
to move to better approximations. In particular incorporating the 
heterogeneous pair approximation \cite{pugliese,gleeson} would be a natural next step,
the more difficult step to motif based expansions \cite{motif} would provide additional 
benefits. 

For example we may use the symbol $A_{k,j,c}$ to denote the density of nodes of 
type $A$, who have $k$ neighbours of type $A$, $j$ neighbours of type $B$ and that are members of 
$c$ triangles. Incorporating the local triangle count increases the complexity of the equation 
system, and leads to higher dimensional PDEs. However, it should be possible to transform and 
analyze these equations along the lines set out here. On the positive side incorporating local motif counts ensures that the approximation remains valid when these motifs are over-expressed in the network. Thus, by using an approximation that directly accounts for
local triangle density we gain the ability to accommodate processes that affect this property (e.g.~triangle closing or breaking). Using such a an approximation thus gives us the ability to consider processes that affect the local clustering coefficient and at the same time gives us the analytical tools to design networks that self-organize to prescribed patterns of clustering.      

Extension of the proposed method to motif-based approximation will be important for bringing the proposed method closer to real world applications. While we do not expect to find many applications which could profit from networks with self-organizing degree distributions, networks with self-organizing motif distributions would be interesting in a number of fields. Let us in particular mention the field of swarm robotics. A number of recent papers \cite{swarming,couzin,huepe_preprint} have demonstrated that the dynamics of swarms can be understood using network models. To accommodate all the resulting processes and the effects of space future extensions of the present method, including accounting for clustering, will be necessary. However, once implemented these extensions could enable the design of desirable collective dynamics in swarms of robotic agents. 

\section*{Data access statement}
Data files for numerical simulations are available at the University of Bristol data.bris Research Data Repository doi:10.5523/bris.18vxedh472sbtz1xiiraxzbf4


%
\ifCLASSOPTIONcompsoc
  \section*{Acknowledgments}
\else
  \section*{Acknowledgment}
\fi
HS was funded by the UK Engineering \& Physical Sciences Research Council (EPSRC) through the Bristol Centre for Complexity Sciences (EP/I013717/1).
\ifCLASSOPTIONcaptionsoff
  \newpage
\fi

\appendices
\section{Heterogeneous approximation for binary adaptive network}\label{appendix:A}
Here we give the ODEs describing the evolution of the binary adaptive network described in Section~\ref{sec:binary}, subject to constant rewiring and constant state-change processes.
We use a heterogeneous active-neighbourhood approximation to track the abundance of nodes $N_{k,l}$, where $N \in \lbrace A,B \rbrace $.
For nodes $A_{k,l}$ we find
\begin{align*}
	\MoveEqLeft
	\dfrac{{\rm d} A_{k,l}}{{\rm d} t} 
		= w_{\rm AB-AA}
		\left[ (l+1)A_{k-1,l+1} - l A_{k,l}  \right.
		\\ & \qquad \left.
		+ \dfrac{\sum_{k',l'} l' A_{k',l'} }{\sum_{k',l'}  A_{k',l'}}
		\left( A_{k-1,l} - A_{k,l} \right) \right]
		\\ &
		+ w_{\rm AB-BB}
		\left[ (l+1)A_{k,l+1} - l A_{k,l} \right]
		\\ & 
		+ \dfrac{w_{\rm AA-AB}}{2}
		\left[ (k+1)A_{k+1,l-1} - k A_{k,l} \right.
		\\ & \qquad \left. 
		+ (k+1) A_{k+1,l} - k A_{k,l} \right] \stepcounter{equation}\tag{\theequation}\label{eq:AODE}
		\\ & 
		+ \dfrac{w_{\rm BB-AB}}{2}
		\left[ \dfrac{ \sum_{k',l'} l' B_{k',l'}}{\sum_{k',l'} A_{k',l'}}
		\left( A_{k,l-1} - A_{k,l} \right) \right]
		\\ &
		+ p_{\rm A-B}
		\left[\left( (k+1)A_{k+1,l-1} - k A_{k,l} \right) - A_{k,l}  \right]
		\\ &
		+ p_{\rm B-A}
		\left[ \left( (l+1)A_{k-1,l+1} - l A_{k,l} \right) + B_{k,l} \right],
\end{align*}
and similarly for $B_{k,l}$ nodes
\begin{align*}
	\MoveEqLeft
	\dfrac{{\rm d} B_{k,l}}{{\rm d} t} 
		= w_{\rm AB-AA}
		\left[ (k+1) B_{k+1,l} - k B_{k,l} \right]
		\\ &
		+ w_{\rm AB-BB}
		\left[ (k+1) B_{k+1,l-1} - k B_{k,l} \right.
		\\ & \qquad \left.
		+ \dfrac{\sum_{k',l'} k' B_{k',l'} }{\sum_{k',l'} B_{k',l'}}
		\left( B_{k,l-1} - B_{k,l} \right) \right]
		\\ &
		+ \dfrac{w_{\rm AA-AB}}{2}
		\left[ \dfrac{\sum_{k',l'} k' A_{k',l'}}{\sum_{k',l'} B_{k',l'}}
		\left(B_{k-1,l} - B_{k,l} \right) \right]
		\\ &
		+ \dfrac{w_{\rm BB-AB}}{2}
		\left[(l+1) B_{k-1,l+1} - l B_{k,l} \right. \stepcounter{equation}\tag{\theequation}\label{eq:BODE}
		\\ & \qquad \left.
		 + (l+1) B_{k,l+1} - l B_{k,l}  \right]
		\\ & 
		+ p_{\rm A-B}
		\left[\left( (k+1) B_{k+1,l-1} - k B_{k,l} \right) + A_{k,l} \right]
		\\ &
		+ p_{\rm B-A}\left[  \left( (l+1)B_{k-1,l+1} - l B_{k,l} \right) - B_{k,l} \right]
\end{align*}
\section{Generating function PDEs for a binary adaptive network}\label{appendix:B}
We convert the infinite-dimensional systems of ODEs \eqref{eq:AODE}--\eqref{eq:BODE}, into two coupled first-order PDEs by introducing two generating functions $G^A(x,y) = \sum_{k,l} A_{k,l} x^k y^l$ and $G^B(x,y) = \sum_{k,l} B_{k,l} x^k y^l$.
We multiply \eqref{eq:AODE} and \eqref{eq:BODE} by $x^k$ and $y^l$ and sum over $k,l \geq 0$.
In the steady state for $G^A$ we find
\begin{align*}
		\MoveEqLeft 0  =
		G^A_x
		\left[ ( y - x) \left( \dfrac{w_{\rm AA-AB}}{2} + p_{\rm A-B}\right)\right.
		 \\ & \qquad \left.
		+ \dfrac{w_{\rm AA-AB}}{2} (1 - x) \right]
		\\ & 
		+ G^A_y
		\left[ (x - y) \left( w_{\rm AB-AA} + p_{\rm B-A} \right) \right. 
		\\ & \left.  \qquad		
		+ w_{\rm AB-BB}(1 - y) \right]  \stepcounter{equation}\tag{\theequation}\label{eq:APDE}
		\\ &
		+ G^A \left[ w_{\rm AB-AA}
		\dfrac{\bar{G}^A_y}{\bar{G}^A}(x - 1) \right.
		\\ & \qquad \left.
		+ \dfrac{w_{\rm BB-AB}}{2} 
		\dfrac{\bar{G}^B_y}{\bar{G}^A}( y - 1 ) - p_{\rm A-B} \right]
		\\ &
		+ p_{\rm B-A}G^B,
\end{align*}
and similarly for $G^B(x,y)$
\begin{align*}
	\MoveEqLeft 0 =
	 G^B_x \left[ (y - x) \left( w_{\rm AB-BB} + p_{\rm A-B} \right)\right.
	 \\ & \qquad \left. 
	  +  w_{\rm AB-AA} (1 - x) \right] 
	 \\ &
	 + G^B_y \left[ (x - y) \left( \frac{w_{\rm BB-AB}}{2} +p_{\rm B-A} \right)\right.	 
	 \\ & \qquad  \left.
	  + \frac{w_{\rm BB-AB}}{2}(1 - y)\right] \stepcounter{equation}\tag{\theequation}\label{eq:BPDE}
	 \\ & 
	  + G^B \left[ w_{\rm AB-BB}
	  \dfrac{\bar{G}^B_{x}}{\bar{G}^B}(y - 1)   \right.
	  \\ & \qquad \left.
	  +  \frac{w_{\rm AA-AB}}{2}
	  \dfrac{\bar{G}^A_x}{\bar{G}^B}(x - 1) - p_{\rm B-A} \right] 
	  \\ &
	  + p_{\rm A-B} G^A,
\end{align*} 
where, for example, $\bar{G}^A = G^A(1,1)$.
\section{Generating function PDEs for a binary adaptive network with degree-dependent state-change rates}\label{appendix:C}
In Section~\ref{sec:binary} we introduce degree-dependent state change processes into the system, while rewiring remains constant. Hence $A$-nodes change state at a rate $\alpha_{k,l}$ and $B$-nodes change state at a rate $\beta_{k,l}$. We therefore introduce two new generating functions $S(x,y) = \sum_{k,l}\alpha_{k,l} A_{k,l}x^k y^l$ and $T(x,y) = \sum_{k,l} \beta_{k,l}B_{k,l} x^k y^l$.
In the steady-state for $G^A$ we find
\begin{subequations}\label{eq:APDE2}
\renewcommand{\theequation}{\theparentequation \roman{equation}}
	\begin{align}
	\MoveEqLeft 0  = G^A_x \left[\dfrac{w_{AA-AB}}{2} ( y - 2 x + 1) \right] \label{eq:APDE2i}
		\\ &
		+ G^A_y \left[ w_{AB-AA} (x - y) + w_{AB-BB}(1 - y) \right] \label{eq:APDE2ii}
		\\ & 
		+ G^A \left[ w_{AB-AA} \dfrac{\bar{G}^A_y}{\bar{G}^A}(x - 1) \right. \notag
		\\ & \qquad \left. 
		+ \dfrac{w_{BB-AB}}{2} \dfrac{\bar{G}^B_y}{\bar{G}^A}( y - 1 ) \right]  \label{eq:APDE2iii}
		\\ & 
		+ \dfrac{\bar{S}_x}{\bar{G}^A_x}(y - x)G^A_x
		+ \dfrac{\bar{T}_x}{\bar{G}^A_y}(x-y)G^A_y \label{eq:APDE2iv}
		- S + T ,
\end{align}
\end{subequations}
and similarly for $G^B$ 
\begin{subequations}\label{eq:BPDE2}
\renewcommand{\theequation}{\theparentequation \roman{equation}}
	\begin{align}
	\MoveEqLeft 0 =
	 G^B_x \left[ w_{AB-BB} (y - x) + w_{AB-AA}(1-x)\right] \label{eq:BPDE2i}
	 \\ &
	 + G^B_y \left[  \frac{w_{BB-AB}}{2}(x - 2 y + 1)  \right]  \label{eq:BPDE2ii}
	 \\ &
	  + \left[ w_{AB-BB} \dfrac{\bar{G}^B_{x}}{\bar{G}^B}(y - 1) \right. \notag
	  \\ & \qquad \left.
	  +  \frac{w_{AA-AB}}{2}\dfrac{\bar{G}^A_x}{\bar{G}^B}(x - 1) \right] G^B \label{eq:BPDE2iii}
	  \\ &
	  + \dfrac{\bar{S}_y}{\bar{G}^B_x}(y - x)G^B_x
	  + \dfrac{\bar{T}_y}{\bar{G}^B_y}(x-y)G^B_y
	   + S - T. \label{eq:BPDE2iv}
	\end{align}
\end{subequations}
Substituting the target distributions $G^A(x,y)$ and $G^B(x,y)$ from section~\ref{sec:binary} into (\ref{eq:APDE2}) and (\ref{eq:BPDE2}) gives two equations in six unknowns.
This shows that the system is still under-determined and we have the freedom to impose additional constraints to arrive at a solution.
Hence here we solve for the rewiring processes and state change processes separately.
We set \eqref{eq:APDE2i} + \eqref{eq:APDE2ii} + \eqref{eq:APDE2iii} and \eqref{eq:BPDE2i} + \eqref{eq:BPDE2ii} + \eqref{eq:BPDE2iii} equal to zero and solve to get a relation between the rewiring processes, while \eqref{eq:APDE2iv}  and \eqref{eq:BPDE2iv} give a relation between the state-change processes.


\bibliographystyle{IEEEtran}
\bibliography{IEEEabrv,triple_jump_references}
\end{document}